\baselineskip=15pt
\font\msbm=msbm10
\magnification=\magstep1
\centerline{\bf{ON p-ADIC FUNCTIONAL INTEGRATION}}
\vskip0.5cm
\centerline{Goran S. DJORDJEVI\'C, ${\rm Branko \  DRAGOVICH}^*$}
\vskip0.5cm
\centerline{{\it Department of Physics, University of Ni\v s,}}
\centerline{{\it P.O.Box 91, 18001 Ni\v s, Yugoslavia}}
\centerline{E-mail: gorandj@junis.ni.ac.yu}
\vskip0.5cm
\centerline{${}^*${\it Institute of Physics, P.O.Box 57, 11001}}
\centerline{{\it Belgrade, Yugoslavia}}
\centerline{E-mail: dragovic@phy.bg.ac.yu}
\vskip1cm

\centerline{{\bf Abstract}}
\vskip0.5cm

{\it p-Adic generalization of the Feynman path integrals in quantum
mechanics is considered. The probability amplitude ${\cal
K}_v(x^{\prime\prime},t^{\prime\prime};x^\prime,t^\prime)$\hfill
$(v = \infty,2,3,\cdots,p,\cdots)$ for a particle in a constant field
is calculated. Path integrals over $\msbm \hbox{Q}_p$ have the same
form as those over $\msbm \hbox{R}$.}

\vskip1cm
1. {\bf Introduction}
\vskip0.5cm

This investigation of p-adic functional integration is motivated by a
successful application of p-adic numbers in many parts of theoretical
and mathematical physics, and particularly in quantum mechanics (for
a review see Refs. 1,2,3). As it is known [4], p-adic numbers may be
obtained by completion of the field of rational numbers $\msbm \hbox{Q}$ with
respect to the p-adic norm (valuation). Recall that the field of real
numbers $\msbm \hbox{R}$ may be regarded as completion of $\msbm \hbox{Q}$ with respect to the
usual absolute value.

In physical measurements only rational numbers can be obtained, and
comparison of theoretical models and experimental results performs
within $\msbm \hbox{Q}$. Since there is a good analysis on $\msbm \hbox{R}$ theoretical models
are traditionally constructed over $\msbm \hbox{R}$ and not over
$\msbm \hbox{Q}$. Only in 1987 the use of
p-adic analysis started to give acceptable physical models.

The fields of p-adic numbers $\msbm \hbox{Q}_p$ (p denotes any prime number) and
$\msbm \hbox{R}$ exhaust all possible number fields which can be obtained by
completion of $\msbm \hbox{Q}$. It is a natural hope that inclusion of p-adic
numbers in the study of physical phenomena provides some additional
clarification of a problem in question. The space of adeles is a
mathematical concept which contains all $\msbm \hbox{Q}_p$ and $\msbm \hbox{R}$, and enables to
treat them simultaneously.

One of the greatest achievements in the use of p-adic numbers in
physics is a formulation of p-adic quantum mechanics [5]. The elements
of the corresponding Hilbert space $L_2(\msbm \hbox{Q}_p)$ are complex-valued
functions of a p-adic argument. It is recently shown [6] that p-adic
quantum mechanics and the ordinary one can be unified in the form of
adelic quantum mechanics, which is quantum mechanics over the space
of adeles.

One of the basic characteristics of quantum mechanics is a
probabilistic description of physical events. As a starting point to
calculate quantum dynamics, instead of the Schr\"odinger
equation, one can use the kernel
${\cal K}(x^{\prime\prime},t^{\prime\prime};x^\prime,t^\prime)$ of the
evolution operator $U(t)$. This
${\cal K}(x^{\prime\prime},t^{\prime\prime};x^\prime,t^\prime)$ is also
called the quantum-mechanical propagator, or the probability
amplitude for a particle to go from a space-time point
$(x^\prime,t^\prime)$ to a space-time point
$(x^{\prime\prime},t^{\prime\prime})$. For the reason of simplicity
we consider only 1+1 - dimensional systems.

Feynman has postulated [7]
${\cal K}(x^{\prime\prime},t^{\prime\prime};x^\prime,t^\prime)$ to be
the path integral
$$
{\cal K}(x^{\prime\prime},t^{\prime\prime};x^\prime,t^\prime) = \int
\exp({2\pi i\over h}S[q]){\cal D}q\ ,\eqno(1.1)
$$
where $S[q] = \int^{t^{\prime\prime}}_{t^{\prime}}L(q,\dot q,t)dt$ is an
action. $L(q,\dot q,t)$ is the Lagrange function and $x^{\prime\prime} =
q(t^{\prime\prime}), \  x^\prime = q(t^\prime)$. The integral in (1.1)
is a symbol which has to realize an intuitive understanding that a
quantum-mechanical particle can propagate from $x^\prime$ to
$x^{\prime\prime}$ using uncountably many paths which connect these
two points. Thus the Feynman path integral means a continual
summation of single amplitudes $\exp({2\pi i\over h}S[q])$ over all paths
$q(t)$ staying between $x^\prime = q(t^\prime)$ and $x^{\prime\prime}
= q(t^{\prime\prime})$. In practical calculations it is the limit
 of an ordinary multiple integral over $N-1$
variables $q_i = q(t_i)$ when $N\to\infty$. Namely, the time interval
$t^{\prime\prime}-t^\prime$ is divided into $N$ equal subintervals and
integration is taken over $q_i$ from $-\infty$ to $+\infty$, but for
a fixed time $t_i$.

The kernel ${\cal K}$ defined in (1.1) must satisfy the following conditions:
\vskip0.5cm
{\settabs 10 \columns
\+\hskip.5cm {\it (i)}&&${\cal K}(x^{\prime\prime},t^{\prime\prime};x^\prime,t^\prime) = \int_R
{\cal K}(x^{\prime\prime},t^{\prime\prime};x,t){\cal K}(x,t;x^\prime,t^\prime)dx$&&&&&&&
(1.2)\cr
\vskip0.5cm
\+\hskip.5cm {\it (ii)}&&$\int_R{\cal K}^*(x^{\prime\prime},t^{\prime\prime};x,t) 
{\cal K}(x^\prime,t^{\prime\prime};x,t)dx = \delta(x^{\prime\prime}-x^\prime)$&&&&&&&
(1.3)\cr
\vskip0.5cm
\+\hskip.5cm {\it (iii)}&&${\cal K}(x^{\prime\prime},t;x^\prime,t) = \delta(x^{\prime\prime}-x^\prime).$&&&&&&&(1.4)\cr}
\vskip0.5cm

For the classical action $\bar
S(x^{\prime\prime},t^{\prime\prime};x^\prime, t^\prime)$, {\it i.e.}
the minimal action with the classical trajectory $\bar q = \bar
q(t)$, which is quadratic in $x^{\prime\prime}$ and $x^\prime$, it
has been shown [8] that in the real case 
$$
{\cal K}(x^{\prime\prime},t^{\prime\prime};x^\prime,t^\prime) = \bigg(
{i\over h}{\partial^2\bar S\over\partial x^{\prime\prime}\partial x^\prime}
\bigg)^{{1\over2}}\exp\big({2\pi i\over h}\bar S(x^{\prime\prime},t^
{\prime\prime};x^\prime,t^\prime)\big)\ ,\eqno(1.5)
$$
where $h$ is the Planck constant.

Feynman invented path integral in 1942, but he published it in 1948
[7]. Since then it has been a subject of permanent interest in
theoretical and mathematical physics, and it is now one of the best
approaches to quantum theory. On the recent progress, present state
and some future prospects one can see Ref. 9.

This paper is devoted to the p-adic generalization of (1.1), {\it i.e.} 
$x^{\prime\prime},x^\prime;$\hfill $t^{\prime\prime},
t^\prime$ are p-adic
variables and ${\cal K}$ is the complex-valued function. Such approach
is suggested in Ref. 5 and its validity is
demonstrated on the
harmonic oscillator [10]. We shall see here that it can be formally
done in the same way, and that it has the same form, as in the real case.
\vskip0.5cm
2. {\bf On p-adic numbers and p-adic analysis}
\vskip0.5cm

We recall here only some basic properties which are relevant to the
rest of the paper.

Any $x\in \msbm \hbox{Q}_p$ can be presented in the form
$$
x = p^\nu(x_0+x_1p+x_2p^2+\cdots)\ ,\quad \nu\in\msbm \hbox{Z}\ ,\eqno(2.1)
$$
where $x_i = 0,1,\cdots,p-1$ are digits. p-Adic norm of any term
$x_ip^{\nu+i}$ in the canonical expansion (2.1) is $\mid
x_ip^{\nu+i}\mid_p = p^{-(\nu+i)}$. Since p-adic norm is the
non-archimedean one, {\it i.e.} $\mid a+b\mid_p\leq\hbox{max}\{\mid
a\mid_p,\mid b\mid_p\}$, it follows that $\mid x\mid_p = p^{-\nu}$.
p-Adic numbers are closely connected to the prime numbers $(p =
2,3,5,\cdots)$ and there are infinitely many $\msbm \hbox{Q}_p$, {\it
i.e.} for every $p$ there is one corresponding $\msbm \hbox{Q}_p$.

There is no natural ordering on $\msbm \hbox{Q}_p$. However one can
introduce a linear order on $\msbm \hbox{Q}_p$ by the following
definition: $x<y$ if $\mid x\mid_p<\mid y\mid_p$ or when 
$\mid x\mid_p = \mid y\mid_p$ there exists such index $m\geq0$ that
digits satisfy $x_0 = y_0, x_1 = y_1, \cdots,x_{m-1} = y_{m-1}\ ,x_m<y_m$.

There are two kinds of analysis over $\msbm \hbox{Q}_p$: the first
one which is connected with map $\varphi : \msbm \hbox{Q}_p\to\msbm
\hbox{Q}_p$, and another one related to map $f:\msbm \hbox{Q}_p\to
\msbm \hbox{C}$, where $\msbm \hbox{C}$ is the field of complex
numbers. Here, we use the both of these analysis. Derivatives of
$\varphi(x)$ are defined as in the real case, but with respect to the
p-adic norm. There is no integral $\int\varphi(x)dx$ in a sense of
the Lebesgue measure [4], but one can introduce $\int^b_a\varphi(x)dx
= \Phi(b)-\Phi(a)$ as a functional of analytic functions
$\varphi(x)$, where $\Phi(x)$ is an antiderative of $\varphi(x)$. In
the case of map $f:\msbm \hbox{Q}_p\to\msbm \hbox{C}$ there is
well-defined Haar integral. We use here the Gauss integral
$$
\int_{\msbm \hbox{Q}_v}{\cal X}_v(ax^2+bx)dx =
\lambda_v(a)\mid2a\mid^{-{1\over2}}_v{\cal
X}_v\big(-{b^2\over4a}\big)\ ,\qquad a\not=0\ ,\eqno(2.2)
$$
where index $v$ denotes real and any of p-adic cases, {\it i.e.} $v =
\infty,2,3,5,\cdots,$ where $\infty$ corresponds to the real case.
${\cal X}_v$ is a multiplicative character: ${\cal X}_\infty(x) =
\exp(-2\pi ix),\, \  {\cal X}_p(x) = \exp(2\pi i\{x\}_p)$ where $\{x\}_p$
is the fractional part of $x\in \msbm \hbox{Q}_p$. Function
$\lambda_v(a)$ is the complex-valued one defined as follows [1]:
$$
\lambda_\infty(a) = {1\over\sqrt2}(1-i\hbox{sign}a)\ ,\eqno(2.3)
$$
$$
\lambda_p(a) = \cases{1,&$\nu = 2k{}{}{}{} \ \  \ \ \ ,\quad p\not=2$,\cr
\big({a_0\over p}\big),&$\nu = 2k+1,\quad p\equiv1(\hbox{mod 4})$,\cr
i\big({a_0\over p}\big)&$\nu = 2k+1,\quad p\equiv3(\hbox{mod 4})$,\cr}\eqno(2.4)
$$
$$
\lambda_2(a) = \cases{{1\over\sqrt2}[1+(-1)^{a_{1}}i],&$\nu = 2k$\ ,\cr
{1\over\sqrt2}(-1)^{a_{1}+a_{2}}[1+(-1)^{a_{1}}i],&$\nu = 2k+1$,\cr}
\eqno(2.5)
$$
where p-adic number $a$ has the form (2.1), $\big({a_0\over p}\big)$
is the Legendre symbol, and $k\in\msbm\hbox{Z}$. We will mainly
use the following properties of $\lambda_v(a)$:
$$
\eqalign{&\lambda_v(0) = 1,\quad \lambda_v(a^2b) = \lambda_v(b),\quad
\lambda_v(a)\lambda_v(b) = \lambda_v(a+b)\lambda_v({1\over a}+{1\over
b}),\cr
&\lambda_v^*(a)\lambda_v(a) = 1\ .\cr}\eqno(2.6)
$$
\vskip0.5cm
3. {\bf p-Adic generalization of the Feynman path integral}
\vskip0.5cm

Let us now consider
$$
{\cal K}_v(x^{\prime\prime},t^{\prime\prime};x^\prime,t^\prime) =
\int{\cal X}_v(-{1\over h}S[q]){\cal D}_vq\eqno(3.1)
$$
as a generalization of (1.1) in ordinary quantum mechanics on all
number fields $\msbm \hbox{Q}_v$ which can be obtained by completion
of $\msbm \hbox{Q}$ (note that $\msbm \hbox{Q}_\infty\equiv\msbm \hbox{R}$).
In (3.1) ${\cal K}_v$ is a complex-valued function of $x^{\prime\prime},
t^{\prime\prime},x^\prime,t^\prime\in\msbm \hbox{Q}_v \ (v =
\infty,2,3, \cdots)$.
\vskip0.5cm
\noindent
DEFINITION. If there exists the limit
$$
\eqalign{&\lim_{N\to\infty}{\cal K}_v^{(N)}(x^{\prime\prime},t^{\prime\prime}
;x^\prime,t^\prime) = \lim_{N\to\infty}A_N\int_{\msbm \hbox{Q}_v}\cdots
\int_{\msbm \hbox{Q}_v}\cr
&{\cal X}_v
\bigg(-{1\over h}\sum^N_{i=1}\bar S(q_i,t_i;q_{i-1},t_{i-1})\bigg)dq_1\cdots
dq_{N-1}\ ,\cr}\eqno(3.2)
$$
where $x^\prime = q_0,\quad x^{\prime\prime} = q_N,\quad t^\prime =
t_0,\quad t^{\prime\prime} = t_N$, and $A_N = A_N(t^{\prime\prime},t^\prime)$ 
is a normalization factor, we say  

$$
{\cal K}_v(x^{\prime\prime},t^{\prime\prime}
;x^\prime,t^\prime) = \lim_{N\to\infty}{\cal K}_v^{(N)}(x^{\prime\prime},
t^{\prime\prime};x^\prime,t^\prime)\eqno(3.3)
$$
is the Feynman path integral.

It is understood that $t_0<t_1<\cdots<t_{N-1}<t_N$ according to
linear order on $\msbm \hbox{Q}_v$ and that $\mid
t_i-t_{i-1}\mid_v\to0$ for every $i = 1,2,\cdots,N$ when
$N\to\infty$. Note that $\bar S$ in (3.2) is the classical action
which corresponds to the classical path $\bar q(t)$. This path $\bar
q(t)$ satisfies the Euler-Lagrange equation
$$
{\partial L\over\partial q}-{d\over dt}{\partial L\over\partial \dot q} =
0\ ,\quad L = L(q,\dot q,t)\eqno(3.4)
$$
and yields the minimum (extreme) of the action $S[q]$.
\vskip0.5cm
4. {\bf Generalized path integral for some physical systems}
\vskip0.5cm

To present time the path integral (3.1) has been derived in the real
case for many physical systems [9]. However, p-adic path integral is
calculated only for the harmonic oscillator [10]. Here we give the
result of calculation for a particle in a constant external field.
For the real case this integral can be found in Ref. 7. For the
reason of simplicity we use units in which $m = h = 1$, where $m$ is
the mass of a particle and $h$ is the Planck constant.

The Lagrange function of a particle with a constant acceleration $a$ is
$$
L(q,\dot q) = {\dot q^2\over2}+aq\ ,\eqno(4.1)
$$
where $q$ denotes position. The corresponding Euler-Lagrange equation
(3.4) is $\ddot q = a$. If we denote $t^\prime = 0, \  t^{\prime\prime}
= T, \  q(0) = q^\prime, \  q(T) = q^{\prime\prime}$ one obtains the
classical path
$$
q(t) = {a\over2}t^2+{q^{\prime\prime}-q^\prime-{a\over2}T^2\over
T}t+q^\prime\ .\eqno(4.2)
$$
For (4.2) one can calculate the classical action
$$
\bar S(q^{\prime\prime},T;q^\prime,0) = \int^T_0\bigg({\dot q^2\over2}+aq\bigg)dt
\eqno(4.3)
$$
and it yields
$$
\bar S(q^{\prime\prime},T;q^\prime,0) = {(q^{\prime\prime}-q^\prime)^2\over2T}
+{a\over2}(q^{\prime\prime}+q^\prime)T-{a^2\over24}T^3\ .\eqno(4.4)
$$
Note that expressions (4.1) to (4.4) are valid for real and all
p-adic cases.

Let us decompose the time interval $T$ into $N$ subintervals $0 =
t_0<t_1<\cdots<t_{N-1}<t_N = T$ and denote $\varepsilon_i =
t_i-t_{i-1}, \ q_i = q(t_i)$ for $i = 1,\cdots,N$. To use definition
(3.2) let us prove the following
\vskip0.5cm
\noindent
PROPOSITION. It holds
$$
\eqalign{&\prod^N_{i=1}\lambda_v(2\varepsilon_i)\mid\varepsilon_i\mid_v^
{-{1\over2}}\int_{\msbm \hbox{Q}_v}\cdots\int_{\msbm \hbox{Q}_v}{\cal
X}_v\bigg[-{1\over2}\sum^N_{i=1}{(q_i-q_{i-1})^2\over\varepsilon_i}-{a\over2}
\sum^N_{i=1}\cr
&(q_i+q_{i-1})\varepsilon_i+
{a^2\over24}\sum^N_{i=1}\varepsilon^3_i\bigg]dq_1\cdots dq_{N-1} = \lambda_v
\bigg(2\sum^N_{i=1}\varepsilon_i\bigg)\mid\sum^N_{i=1}\varepsilon_i\mid_v^
{-{1\over2}}\cr
&{\cal X}_v
\bigg[-{1\over2}{(q_N-q_0)^2\over\sum^N_{i=1}\varepsilon_i}-{a\over2}(q_N+q_0)
\sum^N_{i=1}\varepsilon_i+{a^2\over24}\bigg(\sum^N_{i=1}\varepsilon_i\bigg)^3
\bigg]\ .\cr}\eqno(4.5)
$$

In order to prove (4.5) we use the method of induction, integral
(2.2) and properties (2.6) of $\lambda_v$ functions. First we show
validity of (4.5) for $N = 2$, and then we assume it is valid for $N$
and show that it holds for $N+1$.

Denoting $A_N = \prod^N_{i=1}\lambda_v(2\varepsilon_i)\mid\varepsilon_i
\mid_v^{-{1\over2}}$ we have by  (3.2) that ${\cal K}_v^{(N)}
(x^{\prime\prime},t^{\prime\prime};x^\prime,t^\prime)$ for a particle
of a constant acceleration is equal to the left-hand side of (4.5).
According to the above proposition $\mathop{\lim}\limits_{N\to\infty}
{\cal K}_v^{(N)}
(x^{\prime\prime},t^{\prime\prime};x^\prime,t^\prime)$ is the limit
of the right-hand side of (4.5) and it exists. Recall that $q_N =
q^{\prime\prime}$ for any $N$ and $q_0 = q^\prime$, as well as
$t^{\prime\prime} = T$ and $t^\prime = 0$. Thus we can write down
$$
\eqalign{&{\cal K}_v(x^{\prime\prime},T;x^\prime,0) = \lambda_v(2T)\mid T\mid_v^
{-{1\over2}}\cr
&{\cal X}_v
\bigg[-{1\over2}{(q^{\prime\prime}-q^\prime)^2\over T}-{a\over2}(q^
{\prime\prime}+q^\prime)T+{a^2\over24}T^3\bigg]\ .\cr}\eqno(4.6)
$$

Looking at (4.6) one can see that it has the form
$$
{\cal K}_v(x^{\prime\prime},T;x^\prime,0) = A_v(T){\cal X}_v[-\bar S(x^{\prime
\prime},T;x^\prime,0)]\ .\eqno(4.7)
$$
In the real case $A_\infty(T)$ is usually written as
$(iT)^{-{1\over2}}$ [7], but one can easily show that it is equal to $\lambda
_\infty(2T)\mid T\mid_\infty^{-{1\over2}}$, where $\mid \
\mid_\infty$ denotes the standard absolute value.

To obtain the corresponding result for a free particle it is enough
to put acceleration $a = 0$ in (4.6), {\it i.e.} for a free particle
we get
$$
{\cal K}_v(x^{\prime\prime},T;x^\prime,0) = \lambda_v(2T)\mid T\mid_v^
{-{1\over2}}{\cal X}_v
\bigg[-{1\over2}{(q^{\prime\prime}-q^\prime)^2\over T}\bigg]\ ,\eqno(4.8)
$$
and it is in agreement with the known result in the real case [7].

Using the above procedure we also calculated the Feynman path
integral for the following physical models:

\noindent 
{\it (i)} the de Sitter minisuperspace model [11]
$$
\eqalign{&{\cal K}_v(x^{\prime\prime},T;x^\prime,0) = \lambda_v(-2T)\mid 4T\mid_v^
{-{1\over2}}\cr
&{\cal X}_v
\bigg[{(q^{\prime\prime}-q^\prime)^2\over 8T}
+{\lambda(q^{\prime\prime}+q^\prime)-2\over4}T-{\lambda^2\over24}T^3\bigg]\
,\cr}\eqno(4.9)
$$
{\it (ii)} the harmonic oscillator with a time-dependent frequency [12], [13]
$$
\eqalign{&{\cal K}_v(x^{\prime\prime},t^{\prime\prime};x^\prime,t^\prime) = \lambda_v
\big(2\sin(\gamma^{\prime\prime}-\gamma^\prime)\big)\mid{\sqrt{
\dot\gamma^{\prime\prime}\dot\gamma^\prime}\over\sin(\gamma^{\prime\prime}-
\gamma^\prime)}\mid_v^{{1\over2}}\cr
&{\cal X}_v\bigg[{1\over2}\bigg({\dot s^\prime{x^\prime}^2\over s^\prime}
-{\dot s^{\prime\prime}{x^{\prime\prime}}^2\over s^{\prime\prime}}\bigg)\bigg]
{\cal
X}_v\bigg[-{\dot\gamma^{\prime\prime}{x^{\prime\prime}}^2+\dot\gamma^\prime
{x^\prime}^2\over2\tan(\gamma^{\prime\prime}-\gamma^\prime)}+{x^{\prime\prime}
x^\prime\sqrt{\dot\gamma^{\prime\prime}\dot\gamma^\prime}\over\sin(\gamma^{
\prime\prime}-\gamma^\prime)}\bigg]\ ,\cr}\eqno(4.10)
$$
where $\gamma^\prime = \gamma(t^\prime)\ , \gamma^{\prime\prime} = \gamma(
t^{\prime\prime})$ and $s^\prime = s(t^\prime)\ , s^{\prime\prime} = s(t^{\prime\prime})$.
\vskip0.5cm
5. {\bf Concluding remarks}
\vskip0.5cm

In this paper we defined p-adic generalization of the Feynman path
integral in quantum mechanics. Starting from the definition we calculated it for a
particle in a constant external field. Calculation performed
simultaneously for real and p-adic cases, and all expressions have
the same form with respect to $\msbm\hbox{R}$ and every $\msbm\hbox{Q}_p$.

When classical action $\bar S(x^{\prime\prime},t^{\prime\prime}
;x^\prime,t^\prime)$ in the real case is quadratic in $x^{\prime\prime}$ and
$x^\prime$ there is a general solution [8] which, written in our
notation, reads
$$
{\cal K}_\infty(x^{\prime\prime},t^{\prime\prime};x^\prime,t^\prime)
= \lambda_\infty\bigg(-2{\partial^2\bar S\over\partial x^{\prime\prime}
\partial x^\prime}\bigg)\mid{\partial^2\bar S\over\partial x^{\prime\prime}
\partial x^\prime}\mid_\infty^{{1\over2}}{\cal X}_\infty
\big(-\bar
S(x^{\prime\prime},t^{\prime\prime};x^\prime,t^\prime)\big)\ .
$$
It is natural to expect that the general formula holds
$$
{\cal K}_v(x^{\prime\prime},t^{\prime\prime};x^\prime,t^\prime)
= \lambda_v\bigg(-2{\partial^2\bar S\over\partial x^{\prime\prime}
\partial x^\prime}\bigg)\mid{\partial^2\bar S\over\partial x^{\prime\prime}
\partial x^\prime}\mid_v^{{1\over2}}{\cal X}_v
\big(-\bar
S(x^{\prime\prime},t^{\prime\prime};x^\prime,t^\prime)\big)
$$
whenever $\bar
S(x^{\prime\prime},t^{\prime\prime};x^\prime,t^\prime)$ is quadratic
in $x^{\prime\prime}$ and $x^\prime$. Making the Taylor expansion of
$S$ around its classical (extremal) value $\bar S$ and using the
condition (1.3), we can show that ${\cal K}_v$ has the above form,
but existence of the very prefactor $\lambda_v\bigg(-2{\partial^2\bar 
S\over\partial x^{\prime\prime}\partial x^\prime}\bigg)$ still needs
a rigorous proof.
\vfill\eject
\centerline{{\bf References}}
\vskip0.5cm
\item{[1]} V. S. Vladimirov, I. V. Volovich, E. I. Zelenov, {\it p-Adic
Analysis and Mathematical Physics}, World Scientific, Singapore, 1994.
\item{[2]} L. Brekke, P. G. O. Freund, Physics Reports {\bf 233} (1993),1-66.
\item{[3]} A. Khrennikov, {\it p-Adic Valued Distributions in
Mathematical Physics}, Kluwer Acad. Publishers, Dordrecht, 1994.
\item{[4]} W. H. Schikhof, {\it Ultrametric Calculus: An Introduction
to p-Adic Analysis}, Cambridge Univ. Press, Cambridge, 1984.
\item{[5]} V. S. Vladimirov, I. V. Volovich, Commun. Math. Phys. {\bf
123} (1989), 659-676.
\item{[6]} B. Dragovich, Int. Journal of Modern Phys. {\bf A 10}
(1995), 2349-2365.
\item{[7]} R. P. Feynman, Rev. Mod. Phys. {\bf 20} (1948), 367-387; R.
P. Feynman, A. Hibbs, {\it Quantum Mechanics and Path Integrals},
McGraw Hill, New York, 1965.
\item{[8]} C. Morette, Physical Review {\bf 81} (1951), 848-852.
\item{[9]} Journal Math. Phys. {\bf 36} (5) (1995), a special issue on
functional integration.
\item{[10]} E. I. Zelenov, Journal Math. Phys. {\bf 32} (1991), 147-152.
\item{[11]} J. J. Halliwell, J. Louko, Physical Review {\bf D
39}(1989), 2206-2215.
\item{[12]} D. C. Khandekar, S. V. Lawande, Journal Math. Phys. {\bf
16} (1975), 384-388. 
\item {[13]} G. S. Djordjevic, B. Dragovich, {\it Proc. of Conf. QS96 Minsk}, 
World Scientific, Singapore, 1996.
\vskip0.5cm

\end